\documentclass[prd,preprint,tightenlines,floatfix,
preprintnumbers,nofootinbib,eqsecnum,superscriptaddress]{revtex4}

\usepackage{amsmath,amsfonts,amssymb,amstext,mathrsfs}
\usepackage{amsthm}
\usepackage{mathpazo}

\usepackage[dvips]{graphicx}
\usepackage{epsf,float}
\usepackage{revsymb}

\usepackage{dcolumn}
\usepackage{braket}
\usepackage{color,xcolor}
\usepackage{graphicx}
\usepackage{subfigure}
\usepackage{multirow}
\usepackage{tabularx}
\usepackage{pstricks}
\usepackage[section]{placeins}
\usepackage{booktabs}
\usepackage{array}

\usepackage{hyperref}

\begin{document}

\title{The role of meson exchanges in light-by-light scattering}

\author{Piotr Lebiedowicz}
 \email{Piotr.Lebiedowicz@ifj.edu.pl}
\affiliation{Institute of Nuclear Physics Polish Academy of Sciences, Radzikowskiego 152, PL-31-342 Krak\'ow, Poland}

\author{Antoni Szczurek
\footnote{Also at the Faculty of Mathematics and Natural Sciences,
University of Rzesz\'ow, Pigonia 1, PL-35-310 Rzesz\'ow, Poland.}}
\email{Antoni.Szczurek@ifj.edu.pl}
\affiliation{Institute of Nuclear Physics Polish Academy of Sciences, Radzikowskiego 152, PL-31-342 Krak\'ow, Poland}

\begin{abstract}
We discuss the role of meson exchange mechanisms 
in $\gamma \gamma \to \gamma \gamma$ scattering.
Several pseudoscalar ($\pi^0$, $\eta$, $\eta'(958)$, $\eta_c(1S)$, $\eta_c(2S)$), 
scalar ($f_{0}(500)$, $f_0(980)$, $a_0(980)$, $f_0(1370)$, $\chi_{c0}(1P)$) 
and tensor ($f_2(1270)$, $a_2(1320)$, $f_2'(1525)$, $f_2(1565)$,
$a_2(1700)$) mesons are taken into account.
We consider not only $s$-channel but also for the first time
$t$- and $u$-channel meson exchange amplitudes corrected 
for off-shell effects including vertex form factors.
We find that, depending on not well known vertex form factors, 
the meson exchange amplitudes interfere among themselves and
could interfere with fermion-box amplitudes and 
modify the resulting cross sections.
The meson contributions are shown as a function 
of collision energy as well as angular distributions are presented.
Interesting interference effects separately for light pseudoscalar,
scalar and tensor meson groups are discussed.
The meson exchange contributions may be potentially important 
in the context of a measurement performed recently
in ultraperipheral collisions of heavy ions by the ATLAS collaboration.
The light-by-light interactions could be studied in future
in electron-positron collisions by the Belle II at SuperKEKB accelerator.
\end{abstract}


\maketitle

\section{Introduction}

In the Standard Model the light-by-light scattering
is usually assumed to proceed through lepton, quark and $W$ 
gauge boson loops \cite{Bohm:1994sf,Jikia:1993tc,Bern:2001dg,Bardin:2009gq}. 
The QCD and QED corrections to the fermion-loop contributions 
were discussed in \cite{Bern:2001dg} and their contribution
was found to be very small.
In \cite{KLS2017} our group considered also
double fluctuations of photons into light vector mesons and their
subsequent soft interactions treated in the Regge theory.
The correction turned out to be important at higher center-of-mass energies 
($\sqrt{s} > 2$~GeV) and very small scattering angles.
In \cite{KSS2017} we considered in addition two-gluon exchange contribution.
This mechanism, in contrast to the VDM-Regge one survives to larger 
scattering angles.

In the present paper we shall consider several pseudoscalar 
($J^{PC} = 0^{-+}$), scalar ($J^{PC} = 0^{++}$), and tensor 
($J^{PC} = 2^{++}$) meson exchange contributions 
for the $\gamma \gamma \to \gamma \gamma$ elastic scattering.
The meson-$\gamma$-$\gamma$ vertex functions
depend on the quantum numbers of objects involved
and form factors that are in principle a function of four-momenta of 
both photons and meson. 
So far the $\gamma$-$\gamma$-meson processes 
were investigated at $e^+ e^-$ colliders 
for one or both virtual photons and on-shell $\pi^{0}$ 
\cite{BaBar_TFF_2009,Belle_TFF},
$\eta$, $\eta'$ \cite{BaBar_TFF_2011},
$\eta_{c}$ \cite{BaBar_TFF_2010} mesons
and also from recent measurements by the Belle collaboration \cite{Masuda:2015yoh} 
for $f_{0}(980)$ and $f_{2}(1270)$ mesons.
Recently, in \cite{Danilkin} the authors 
formulated light-by-light scattering sum rules,
that lead to relations between the $\gamma^{*} \gamma$ transition
form factors for $C$-even scalar, pseudoscalar, axial-vector
and tensor mesons when assuming sum rule saturation.

The situation for $\gamma \gamma \to \gamma \gamma$ elastic scattering
is different as here photons are on mass shell and meson is off-shell.
We do not know very precisely the dependence of the form factor for off-shell
mesons on their virtuality. 
For the $s$-channel we have time-like meson
while for $t$- and $u$-channels space-like one 
(see Fig.~\ref{fig:diagram_2to2}).
We shall discuss here how the results depend on the form factors.
In the following we shall parametrise it 
in either exponential or monopole type.
We expect that for the $t$ and $u$ diagrams
the corresponding form factors can be rather hard 
(larger cut-off parameter) compared to purely
hadronic processes as they describe 
the coupling of space-like meson to semi point-like objects (photons).

\section{Meson exchange contributions}
\begin{figure}
\includegraphics[width=0.33\textwidth]{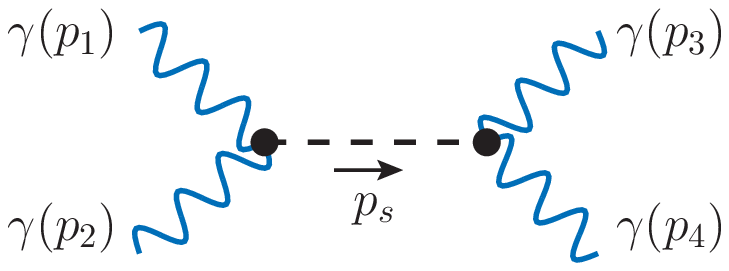}
\includegraphics[width=0.31\textwidth]{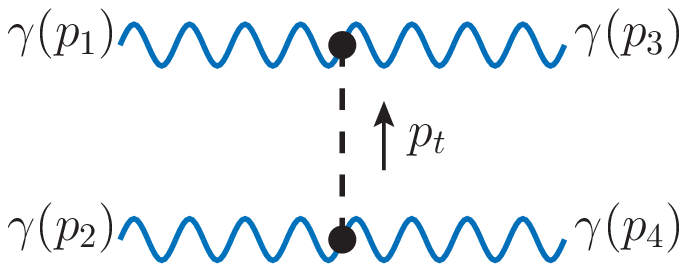}
\includegraphics[width=0.31\textwidth]{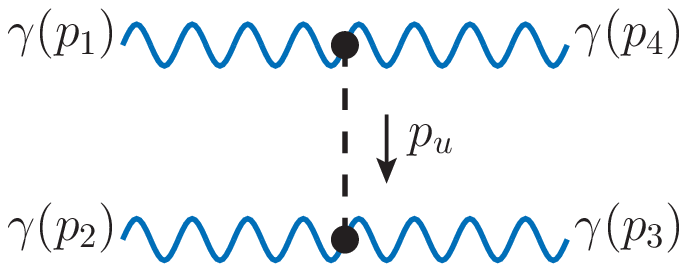}
\caption{\label{fig:diagram_2to2}
Diagrams for light-by-light scattering via a time-like ($s$-channel) 
and a space-like ($t$-channel and $u$-channel) meson exchanges.
}
\end{figure}

Below we will discuss the expressions 
relevant to pseudoscalar ($M_{PS}$), 
scalar ($M_{S}$), and tensor ($M_{T}$)
as well as spin-4 $f_4(2050)$ meson exchange contributions 
for the reaction 
\begin{eqnarray}
\gamma(p_{1},\lambda_{1}) + \gamma(p_{2},\lambda_{2}) \to 
\gamma(p_{3},\lambda_{3}) + \gamma(p_{4},\lambda_{4})
\label{2to2_reaction}
\end{eqnarray}
with the four-momenta $p_i$ $(i = 1, ..., 4)$ and the photon helicities,
$\lambda_{i} \in \{1,-1 \}$, indicated in brackets.
Schematic diagrams for the the process (\ref{2to2_reaction}) 
are given in Fig.~\ref{fig:diagram_2to2}.

The differential cross section for the angular distribution
for the reaction (\ref{2to2_reaction}) 
is given by
\begin{eqnarray}
\frac{d\sigma}{d\cos\theta} =
\frac{1}{32 \pi s} \,
\frac{1}{2} \frac{1}{4} \sum_{\lambda_{1}, \lambda_{2},\lambda_{3}, \lambda_{4}}
|{\cal M}_{\lambda_{1} \lambda_{2} \to \lambda_{3} \lambda_{4}}|^{2} \,,
\label{xsecttion_diff_2to2}
\end{eqnarray}
where the explicit factor $1/2$ takes into account identity of photons.
The kinematical variables used in the present paper are
\begin{eqnarray}
&&s = (p_{1} + p_{2})^{2} = (p_{3} + p_{4})^{2} \,, \nonumber \\
&&t = (p_{1} - p_{3})^{2} = (p_{2} - p_{4})^{2}\,,\nonumber \\
&&u = (p_{2} - p_{3})^{2} = (p_{1} - p_{4})^{2}\,,\nonumber \\
&&p_{s} = p_{1} + p_{2} = p_{3} + p_{4}\,,\nonumber \\
&&p_{t} = p_{2} - p_{4} = p_{3} - p_{1}\,,\nonumber \\
&&p_{u} = p_{1} - p_{4} = p_{3} - p_{2}\,,\nonumber \\
&&p_{s}^{2} = s\,,\; p_{t}^{2} = t\,,\; p_{u}^{2} = u\,.
\label{2to2_kinematic}
\end{eqnarray}

The amplitude for the reaction (\ref{2to2_reaction})
with the meson exchanges is written as
\begin{eqnarray}
{\cal M}_{\lambda_{1}\lambda_{2} \to \lambda_{3}\lambda_{4}}
=&&\sum_{M_{PS} = \pi^{0}, \eta, \eta'(958), \eta_{c}(1S),\eta_{c}(2S)} 
{\cal M}_{\lambda_{1}\lambda_{2} \to \lambda_{3}\lambda_{4}}^{(M_{PS})} 
\nonumber \\
&+&
\sum_{M_{S} = f_{0}(500), f_0(980), a_0(980), f_0(1370), \chi_{c0}(1P)} 
{\cal M}_{\lambda_{1}\lambda_{2} \to \lambda_{3}\lambda_{4}}^{(M_{S})} 
\nonumber \\
&+&\sum_{M_{T} = f_2(1270), a_0(1320), f_{2}'(1525)}
{\cal M}_{\lambda_{1}\lambda_{2} \to \lambda_{3}\lambda_{4}}^{(M_{T})} 
\,.
\label{full_amplitude}
\end{eqnarray}

In Table~\ref{table:PDG} we have collected possible potential resonances
that may contribute to the process (\ref{2to2_reaction}).
The contribution of axial-vector mesons vanishes for on-shell photons due
to the Landau-Yang theorem \cite{LY}.
The two-photon branching fractions for the resonances are
relatively well known and were measured 
in recent years by the Belle and BaBar collaborations.
{\small
\begin{table}[!h]
\caption{A list of mesons that contribute to 
the $\gamma\gamma \to \gamma\gamma$ scattering.
The meson masses, their total widths $\Gamma_{M}$ and
branching fractions, ${\cal B}(M \to \gamma \gamma) = \Gamma(M \to \gamma \gamma)/\Gamma_{M}$,
are taken from PDG~\cite{PDG}.}
\begin{tabular}{|c|l|l|l|l|}
\hline
Meson			& $m_M$ (MeV) 	& $\Gamma_{M}$ (MeV)		&  ${\cal B}(M \to \gamma \gamma)$ & $\Gamma(M \to \gamma\gamma)$ (keV)\\ \hline
$\pi^{0}$		& $134.9766 \pm 0.0006$ 		&  		&  $(98.823 \pm 0.034) \times 10^{-2}$ &
$7.8\times 10^{-3}$ \\ 
$\eta$		& $547.862 \pm 0.017$ 		& $(1.31 \pm 0.05)\times 10^{-3}$ &  $(39.41 \pm 0.20) \times 10^{-2}$ &0.516\\ 
$\eta'(958)$		& $957.78 \pm 0.06$ 		& $0.197 \pm 0.009$ 		&  $(2.21 \pm 0.08) \times 10^{-2}$ &4.354\\ 
$\eta_c(1S)$		& $2 983.4 \pm 0.5$ 		& $31.8\pm 0.8$ 		&  $(1.59 \pm 0.13) \times 10^{-4}$ &5.056\\ 
$\eta_c(2S)$		& $3 639.2 \pm 1.2$ 		&
$11.3^{+3.2}_{-2.9}$ &$(1.9 \pm 1.3) \times 10^{-4}$ &2.147\\
\hline
$f_{0}(500)$        &  $400-550$        &  $400-700$        &  $\mathrm{seen}$      &  $2.05 \pm 0.21$ \cite{Dai_Pennington:2014}      \\
$f_{0}(980)$      &  $990 \pm 20$      &   $10 - 100$       &   $\mathrm{seen}$      &  $0.32 \pm 0.05$ \cite{Dai_Pennington:2014}      \\
$a_{0}(980)$      &  $980 \pm 20$      &   $50 - 100$       &   $\mathrm{seen}$     &   $0.30 \pm 0.10$ \cite{Amsler98}      \\
$f_{0}(1370)$      & $1200-1500$       &  $200-500$        &  $\mathrm{seen}$      & $4.0 \pm 1.9$ \cite{Dai_Pennington:2014}      \\
$\chi_{c0}(1P)$	& $3 414.75 \pm 0.31$		& $10.5\pm0.6$ 		& $(2.23 \pm 0.13) \times 10^{-4}$ &2.342\\ \hline
$f_2(1270)$	&	$1275.5 \pm 0.8$ 			& $186.7^{+2.2}_{-2.5}$	&  $(1.42 \pm 0.24) \times 10^{-5}$ &
2.651; $3.14 \pm 0.20$ \cite{Pennington:2008xd}\\
$a_2(1320)$	&	$1318.3^{+0.5}_{-0.6}$ 			& $107 \pm 5$	&  $(9.4 \pm 0.7) \times 10^{-6}$ &
$1.00 \pm 0.06$
\\
$f'_2(1525)$	&	$1525 \pm 5$ 			& $73^{+6}_{-5}$	&  $(1.10 \pm 0.14) \times 10^{-6}$ & 
$0.081 \pm 0.009$; $0.13 \pm 0.03$ \cite{Shchegelsky} \\
$f_2(1565)$	&	$1562 \pm 13$ 			& $134 \pm 8$	&  & 
$0.70 \pm 0.14$ \cite{Shchegelsky} \\
$a_2(1700)$	&	$1732 \pm 16$ 			& $194 \pm 40$	&  & 
$0.30 \pm 0.05$ \cite{Shchegelsky} \\
\hline
$f_4(2050)$     &   $2018 \pm 11$    & $237 \pm 18$
&        &    0.7 \cite{KS2013}    \\          
\hline
\end{tabular}
\label{table:PDG}
\end{table}
}
\subsection{Pseudoscalar meson exchanges}

The amplitude for the pseudoscalar meson exchange
is written as
\begin{eqnarray}
i{\cal M}_{\lambda_{1}\lambda_{2} \to \lambda_{3}\lambda_{4}}^{(M_{PS})}
&=&
(\epsilon_{3}^{\mu_{3}})^{*}\,
i\Gamma^{(M_{PS} \gamma \gamma)}_{\mu_{3} \mu_{4}}(p_{3},p_{4}) \,
(\epsilon_{4}^{\mu_{4}})^{*}\,
i\Delta^{(M_{PS})}(p_{s})\,
\epsilon_{1}^{\mu_{1}}\,
i\Gamma^{(M_{PS} \gamma \gamma)}_{\mu_{1} \mu_{2}}(p_{1},p_{2}) \,
\epsilon_{2}^{\mu_{2}} \nonumber \\
&&+
(\epsilon_{3}^{\mu_{3}})^{*}\,
i\Gamma^{(M_{PS} \gamma \gamma)}_{\mu_{3} \mu_{1}}(-p_{3},p_{1}) \,
\epsilon_{1}^{\mu_{1}}\,
i\Delta^{(M_{PS})}(p_{t})\,
(\epsilon_{4}^{\mu_{4}})^{*}\,
i\Gamma^{(M_{PS} \gamma \gamma)}_{\mu_{4} \mu_{2}}(p_{4},p_{2})\,
\epsilon_{2}^{\mu_{2}}
\nonumber\\
&&+
(\epsilon_{4}^{\mu_{4}})^{*}\,
i\Gamma^{(M_{PS} \gamma \gamma)}_{\mu_{4} \mu_{1}}(p_{4},p_{1}) \,
\epsilon_{1}^{\mu_{1}}\,
i\Delta^{(M_{PS})}(p_{u})\,
(\epsilon_{3}^{\mu_{3}})^{*}\,
i\Gamma^{(M_{PS} \gamma \gamma)}_{\mu_{3} \mu_{2}}(-p_{3},p_{2})\,
\epsilon_{2}^{\mu_{2}}\,,
\nonumber\\
\label{amplitude_2to2_PS}
\end{eqnarray}
where $\epsilon_{i}^{\mu_{i}}$ are the polarisation vectors of the photons
with the helicities $\lambda_{i}$.

The propagator and $M_{PS} \gamma \gamma$ vertex function 
for the pseudoscalar meson with mass $m_{M_{PS}}$ are: 
\begin{eqnarray}
&&i\Delta^{(M_{PS})}(k) = \frac{i}{k^{2} - m_{M_{PS}}^2 + i m_{M_{PS}} \Gamma_{M_{PS}}}\,,
\label{propagator_PS}\\
&&i\Gamma^{(M_{PS} \gamma \gamma)}_{\mu \nu}(k_{1},k_{2})
= -i e^{2} \,
\varepsilon_{\mu \nu \kappa \lambda} k_{1}^{\kappa} k_{2}^{\lambda} \,
F_{\gamma^{*} \gamma^{*} \to M_{PS}}(0,0)\, F^{(M_{PS} \gamma \gamma)}(k^{2})\,,
\label{vertex_gamgamPS}
\end{eqnarray}
where 
$k^2 = p_{s}^2$, $p_{t}^2$, $p_{u}^2$ for the $s$, $t$, $u$ diagrams, respectively.
The transition form factor $F_{\gamma^{*} \gamma^{*} \to M_{PS}}(0,0)$
is related to the two-photon decay width as
\begin{eqnarray}
F_{\gamma^{*} \gamma^{*} \to M_{PS}}(0,0) 
= \frac{1}{4 \pi^{2} f_{M_{PS}}}
= \frac{1}{4 \pi \alpha_{em}} 
\sqrt{\frac{64 \pi \Gamma(M_{PS} \to \gamma \gamma)}{m_{M_{PS}}^{3}}}\,.
\label{vertex_TFF}
\end{eqnarray}
Above $\alpha_{em} = e^{2}/(4 \pi)$ and $f_{M_{PS}}$ is a decay constant of psudoscalar meson.
We have, from Eq.~(\ref{vertex_TFF}), 
$f_{M_{PS}}$ = 0.092, 0.093, 0.074, 0.375, 0.776~GeV 
for $\pi^{0}$, $\eta$, $\eta'$, $\eta_{c}(1S)$, $\eta_{c}(2S)$ mesons, respectively.
The $f_{M_{PS}}$ can be tried to be calculated within quark model.
For $\eta$ and $\eta'$ this requires inclusion of $\eta$-$\eta'$ mixing 
(see e.g.~\cite{DGH1992}).
The two-photon decay of heavy quarkonium $\eta_{c}$ states
was discussed, e.g., in \cite{LP_etac}.

In our vertices (\ref{vertex_gamgamPS}) 
the photons are on mass shell and the meson is off-shell.
Therefore in calculations we use off-shell meson form factors
normalized to $F^{(M \gamma \gamma)}(m_{M}^{2}) = 1$.
These form factors can be parametrised, e.g., by the exponential form 
for the $t$- and $u$-channel meson exchanges
\begin{eqnarray}
F^{(M \gamma \gamma)}(t/u) = \exp\left(\frac{t/u - m_{M}^{2}}{\Lambda_{exp}^{2}}\right)\,,
\label{exp_ff}
\end{eqnarray}
while for the $s$-channel meson exchange we assume 
\begin{eqnarray}
F^{(M \gamma \gamma)}(s) = \exp\left(-\frac{(s - m_{M}^{2})^{2}}{\Lambda_{exp}^{4}}\right)\,,
\label{exp_ff_s}
\end{eqnarray}
where $\Lambda_{exp}$ is, in principle, a free parameter.
In our calculation for pseudoscalar mesons we take $\Lambda_{exp} = 2$~GeV.

%
%

\subsection{Scalar meson exchanges}

In our calculations here we consider four light scalar resonances: 
$f_{0}(500)$, $f_0(980)$, $a_0(980)$ and $f_{0}(1370)$.
The first two states have been seen 
in $\gamma \gamma \to \pi \pi$ reactions \cite{Belle_pipi},
while the $a_{0}(980)$ resonance was observed 
in the $\gamma \gamma \to \pi^0 \eta$ reaction \cite{Belle_pieta}.
The situation for light scalar mesons was discussed, e.g., in \cite{AS_first}.
While the direct coupling is expected to be rather small, 
the rescattering of mesons ($\pi \pi$, $K K$ or $\pi \eta$) leads to 
two-photon widths of the order of a fraction of keV or even larger.
The partial decay widths were estimated, e.g., 
in \cite{AABN1999,Dai_Pennington:2014} and are collected in Table~\ref{table:PDG}.
Note that the masses and widths of low-mass resonances 
are known with a large uncertainties \cite{PDG}; 
see also analysis in \cite{ALEPH} for the $f_{0}(1500)$ and $f_{0}(1710)$ states.
The two-photon decay of heavy quarkonium states ($\chi_{c0,2}$ and $\chi_{b0,2}$)
was discussed, e.g., in \cite{LP_chic, Danilkin_chic}.

%
%

For the scalar meson $M_{S}$ exchange the amplitude
is similar as (\ref{amplitude_2to2_PS})
with $\Delta^{(M_{PS})}$ and $\Gamma^{(M_{PS} \gamma \gamma)}_{\mu \nu}$
replaced by $\Delta^{(M_{S})}$ and $\Gamma^{(M_{S} \gamma \gamma)}_{\mu \nu}$, respectively.
The corresponding vertex can be written as:
\begin{equation}
i\Gamma^{(M_{S} \gamma \gamma)}_{\mu \nu}(k_1,k_2) = \frac{e^{2}}{2m_{M_{S}}} 
\left( k^2 g_{\mu \nu} - 2\, k_{1 \nu} k_{2 \mu} \right) \,
F_{\gamma^{*} \gamma^{*} \to M_{S}}(0,0)\, F^{(M_{S} \gamma \gamma)}(k^{2})\,,
\label{scalar_vertex}
\end{equation}
where
\begin{eqnarray}
F_{\gamma^{*} \gamma^{*} \to M_{S}}(0,0) 
= \frac{1}{4 \pi \alpha_{em}} 
\sqrt{\frac{64 \pi \Gamma(M_{S} \to \gamma \gamma)}{m_{M_{S}}}}\,.
\label{vertex_TFF_S}
\end{eqnarray}
%

It is easy to check that
\begin{eqnarray}
k_{1 \mu} \Gamma^{\mu \nu}(k_1,k_2) = 0\,, \quad
k_{2 \nu} \Gamma^{\mu \nu}(k_1,k_2) = 0\,.
\label{scalar_gauge_invariance}
\end{eqnarray}
Thus our vertices and as a consequence our $s$-, $t$- and $u$-channel
amplitudes are gauge invariant.

For light scalar $f_0(500)$, $f_0(980)$, $a_0(980)$ and $f_{0}(1370)$ meson exchanges
we take a smaller value of $\Lambda_{exp} = 1$~GeV.
\footnote{This is related to the fact 
that the $\gamma \gamma$ coupling to the scalar mesons is dominantly not to quark-antiquark system
but rather to mesonic loops.}
Their contribution will be shown in final summary plots.

\subsection{Tensor meson exchanges}

The tensor meson exchanges are much more complicated.
The $f_2(1270)$ and $a_2(1320)$ couplings to two photons 
were studied in detail, e.g., in \cite{EMN14} (see section 5).
Using the formalism from \cite{EMN14} 
the amplitude for the tensor meson $M_{T}$ exchange can be written as
\begin{eqnarray}
i{\cal M}_{\lambda_{1}\lambda_{2} \to \lambda_{3}\lambda_{4}}^{(M_{T})}
&=&
(\epsilon_{3}^{\mu_{3}})^{*}\,
i\Gamma^{(M_{T} \gamma \gamma)}_{\mu_{3} \mu_{4} \nu_{3} \nu_{4}}(p_{3},p_{4}) \,
(\epsilon_{4}^{\mu_{4}})^{*}\,
i\Delta^{(M_{T})\, \nu_{3} \nu_{4}, \nu_{1} \nu_{2}}(p_{s})\,
\epsilon_{1}^{\mu_{1}}\,
i\Gamma^{(M_{T} \gamma \gamma)}_{\mu_{1} \mu_{2} \nu_{1} \nu_{2}}(p_{1},p_{2}) \,
\epsilon_{2}^{\mu_{2}}
\nonumber\\
&&+
(\epsilon_{4}^{\mu_{4}})^{*}\,
i\Gamma^{(M_{T} \gamma \gamma)}_{\mu_{4} \mu_{1} \nu_{4} \nu_{1}}(p_{4},p_{1}) \,
\epsilon_{1}^{\mu_{1}}\,
i\Delta^{(M_{T})\, \nu_{4} \nu_{1}, \nu_{3} \nu_{2}}(p_{u})\,
(\epsilon_{3}^{\mu_{3}})^{*}\,
i\Gamma^{(M_{T} \gamma \gamma)}_{\mu_{3} \mu_{2} \nu_{3} \nu_{2}}(p_{3},p_{2})\,
\epsilon_{2}^{\mu_{2}}
\nonumber\\
&&+
(\epsilon_{3}^{\mu_{3}})^{*}\,
i\Gamma^{(M_{T} \gamma \gamma)}_{\mu_{3} \mu_{1} \nu_{3} \nu_{1}}(p_{3},p_{1}) \,
\epsilon_{1}^{\mu_{1}}\,
i\Delta^{(M_{T})\, \nu_{3} \nu_{1}, \nu_{4} \nu_{2}}(p_{t})\,
(\epsilon_{4}^{\mu_{4}})^{*}\,
i\Gamma^{(M_{T} \gamma \gamma)}_{\mu_{4} \mu_{2} \nu_{4} \nu_{2}}(p_{4},p_{2})\,
\epsilon_{2}^{\mu_{2}}\,.
\nonumber\\
\label{amplitude_2to2_T}
\end{eqnarray}

The $f_{2}(1270) \gamma \gamma$ vertex is given as
(see formulae (3.39), (3.40) and section 5.3 of \cite{EMN14})
\begin{equation}
\begin{split}
i \Gamma_{\mu\nu\kappa\lambda}^{(f_2 \gamma \gamma)} (k_1,k_2) 
= i \left[
2 a_{f_2 \gamma\gamma} \Gamma_{\mu\nu\kappa\lambda}^{(0)}(k_1,k_2) 
- b_{f_2 \gamma\gamma} \Gamma_{\mu\nu\kappa\lambda}^{(2)}(k_1,k_2) 
\right] F^{(f_2 \gamma \gamma)} (k^2) \,.
\end{split}
\label{tensor_functions}
\end{equation}
Here the so-called helicity-0 and helicity-2 $f_{2} \to \gamma \gamma$
amplitudes are parametrised by the $a$ and $b$ constants, respectively.
Two rank-four tensor functions, $\Gamma_{\mu\nu\kappa\lambda}^{(0)}$ and $\Gamma_{\mu\nu\kappa\lambda}^{(2)}$,
are defined by (3.18) and (3.19) in \cite{EMN14}.
In our calculation for the tensor mesons ($q \bar{q}$ states) 
we take $\Lambda_{exp} = 2$~GeV in form factors given by (\ref{exp_ff}) and (\ref{exp_ff_s}).
We adopt also the numerical values of the $a_{2}(1320) \gamma \gamma$ coupling constants
discussed in sections~5.4 and 7.2 of \cite{EMN14}.

For a better analysis we shall use model for the $f_{2}$ propagator 
considered in \cite{EMN14}; see Appendix~A there.
The Breit-Wigner formula is used here
\begin{eqnarray}
i\Delta_{\kappa \lambda, \alpha \beta}^{(f_{2})}(k)&=&
\frac{i}{k^{2}-m_{f_{2}}^2+i m_{f_{2}} \Gamma_{f_{2}}}
\left[ 
\frac{1}{2} 
( \hat{g}_{\kappa \alpha} \hat{g}_{\lambda \beta}  + \hat{g}_{\kappa \beta} \hat{g}_{\lambda \alpha} )
-\frac{1}{3} 
\hat{g}_{\kappa \lambda} \hat{g}_{\alpha \beta}
\right] \,, 
\label{prop_f2}
\end{eqnarray}
where $\hat{g}_{\mu \nu} = -g_{\mu \nu} + k_{ \mu} k_{ \nu} / k^2$.

In our work we consider also the $f_{2}'(1525)$ meson contribution.
We assume the same ratio of the two-photon decay widths
of the specific helicity to the total two-photon decay width
as for the $f_{2}(1270)$ meson.
We have
\begin{eqnarray}
a_{f_{2}'(1525) \gamma \gamma} = \pm \frac{e^{2}}{4\pi} 0.12\, \mathrm{GeV}^{-3}\,, \quad
b_{f_{2}'(1525) \gamma \gamma} = \pm \frac{e^{2}}{4\pi} 0.30\, \mathrm{GeV}^{-1}\,.
\label{tensorial_couplings_f2p}
\end{eqnarray}

A contribution of heavier tensor-meson states are also possible, see Table~\ref{table:PDG}.
Recently in \cite{Danilkin} a role of $f_{2}(1565)$ and $a_{2}(1700)$
mesons was discussed in the context of transition form factors.
Here we also take into account these two states.
In order to estimate the strength of couplings
we assume only helicity-2 contributions.
From Eq.~(5.28) of \cite{EMN14} and with the parameters of Table~\ref{table:PDG} we get
\begin{eqnarray}
b_{f_{2}(1565) \gamma \gamma} = \pm \frac{e^{2}}{4\pi} 0.93\, \mathrm{GeV}^{-1}\,, \quad
b_{a_{2}(1700) \gamma \gamma} = \pm \frac{e^{2}}{4\pi} 0.52\, \mathrm{GeV}^{-1}\,.
\label{tensorial_couplings_f2a2}
\end{eqnarray}
In calculation we choose the positive values of above coupling constants.

The $t/u$-channel exchanges of spin-2 particles leads to a growing
of the cross section as a function of $\sqrt{s}$.
In practice the rapid growth appears quickly above the $s$-channel resonance.
No clear unitarization procedure is known to as.
From phenomenology we know that the exchange of the meson
must be replaced by the exchange of the $f_{2}$ reggeon \cite{EMN14}.
In the following we shall therefore omit the $t$- and $u$-channel 
meson exchange contributions.

\subsection{$f_{4}(2050)$ meson exchange}

The $f_{4}(2050)$ resonance was identified 
in $\gamma \gamma \to \pi \pi$ processes \cite{KS2013}.
However, the detailed tensorial coupling for 
$f_{4}(2050) \to \gamma \gamma$ is not known, 
so as a consequence we do not know corresponding angular distributions. 
We shall therefore neglect $t$- and $u$-channel contributions.
In the following we use a simple Breit-Wigner resonance form
for a spin-$J$ resonance $R$:
\footnote{In the general case
the initial particles have spins $s_{1}$ and $s_{2}$.
In our case, for a real photons, a factors $(2 s_{i} + 1)$ are replaced by 2.}
\begin{eqnarray}
\sigma_{\gamma \gamma \to R \to \gamma \gamma}(s)
&=&
\frac{4 (2 J + 1)}{(2 s_1 + 1)(2 s_2 + 1)} 
\frac{4 \pi (s/m_{R}^{2})}
{(s - m_{R}^{2})^{2} + m_{R}^{2} \Gamma_{R}^{2}}  
\left( \sqrt{2} \times \Gamma(R \to \gamma \gamma) \right)^{2} \times \left(F(s)\right)^{4}\,,
\nonumber \\
\label{sigma_f4}
\end{eqnarray}
which represents only $s$-channel contribution to the total cross section. 
%
%
The resonance form reproduces result of our diagrammatic calculation
for lower-spin resonances at $\sqrt{s} \simeq m_{R}$.
In (\ref{sigma_f4}) the resonance width $\Gamma_R$ is a constant. 
An improved description of resonance shapes can be obtained 
when the width is made $s$-dependent.

\section{Results}

In this section we present first numerical results for meson exchange contributions
to the $\gamma \gamma \to \gamma \gamma$ scattering. 
We shall include not only $s$-channel resonant contributions 
but also $t$- and $u$-channel exchanges, 
not discussed so far in the literature.


We start with contributions of the pseudoscalar mesons
that were observed in two-photon processes
in the $e^+ e^-$ collisions (Belle, BaBar).
In the left panel of Fig.~2
we show, for example, separate contributions
of $s$, $t$ and $u$ channels and their coherent sum for $\eta'(958)$ meson.
The $t$ and $u$ channels play a role only for $\sqrt{s} > 2$~GeV.
In the right panel we show separate contributions of $\pi^0$, $\eta$ and
$\eta'$ exchanges as well as their sum.
A large interference effects can be seen.
\begin{figure}
\includegraphics[width=8cm]{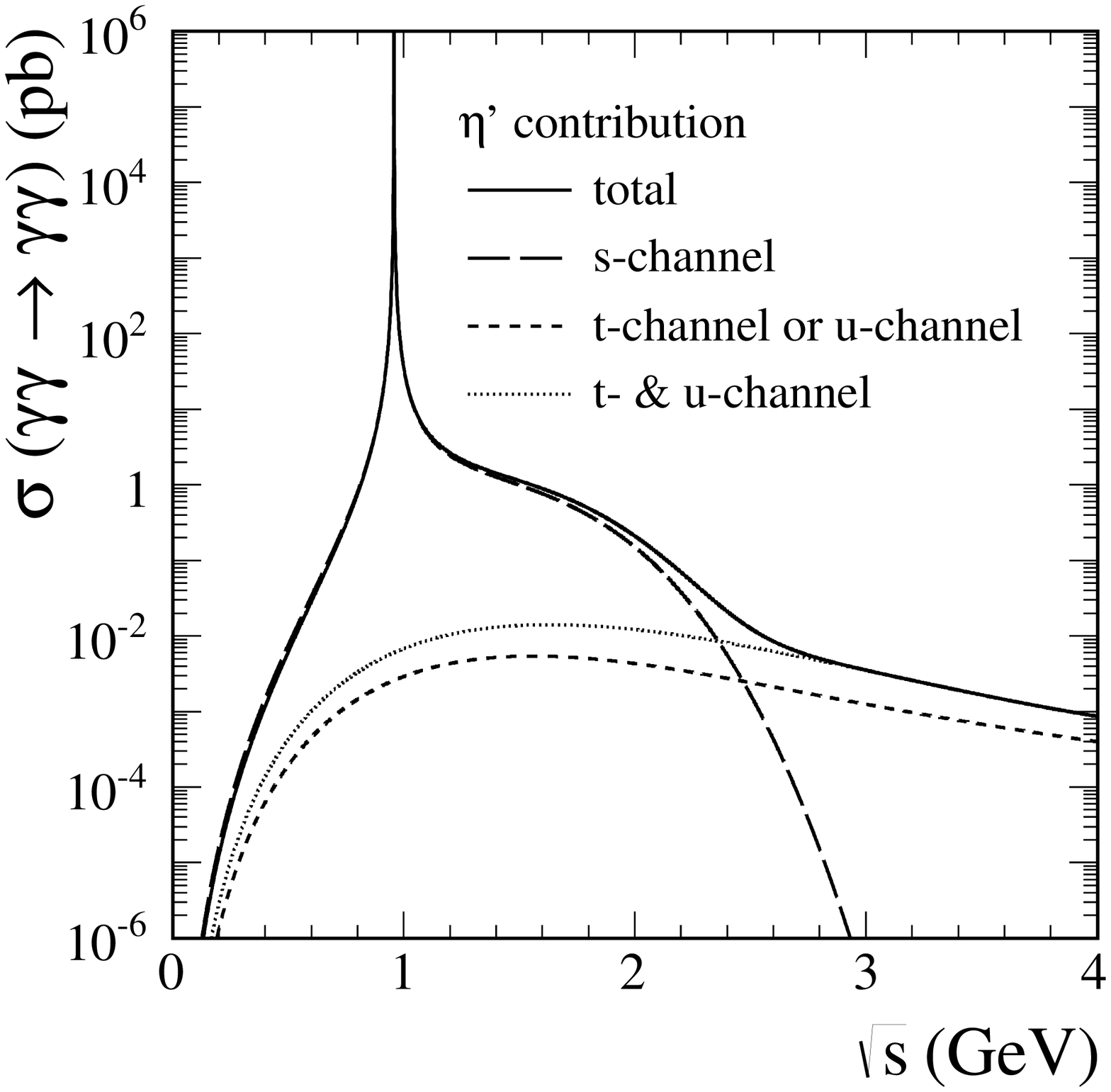}        
\includegraphics[width=8cm]{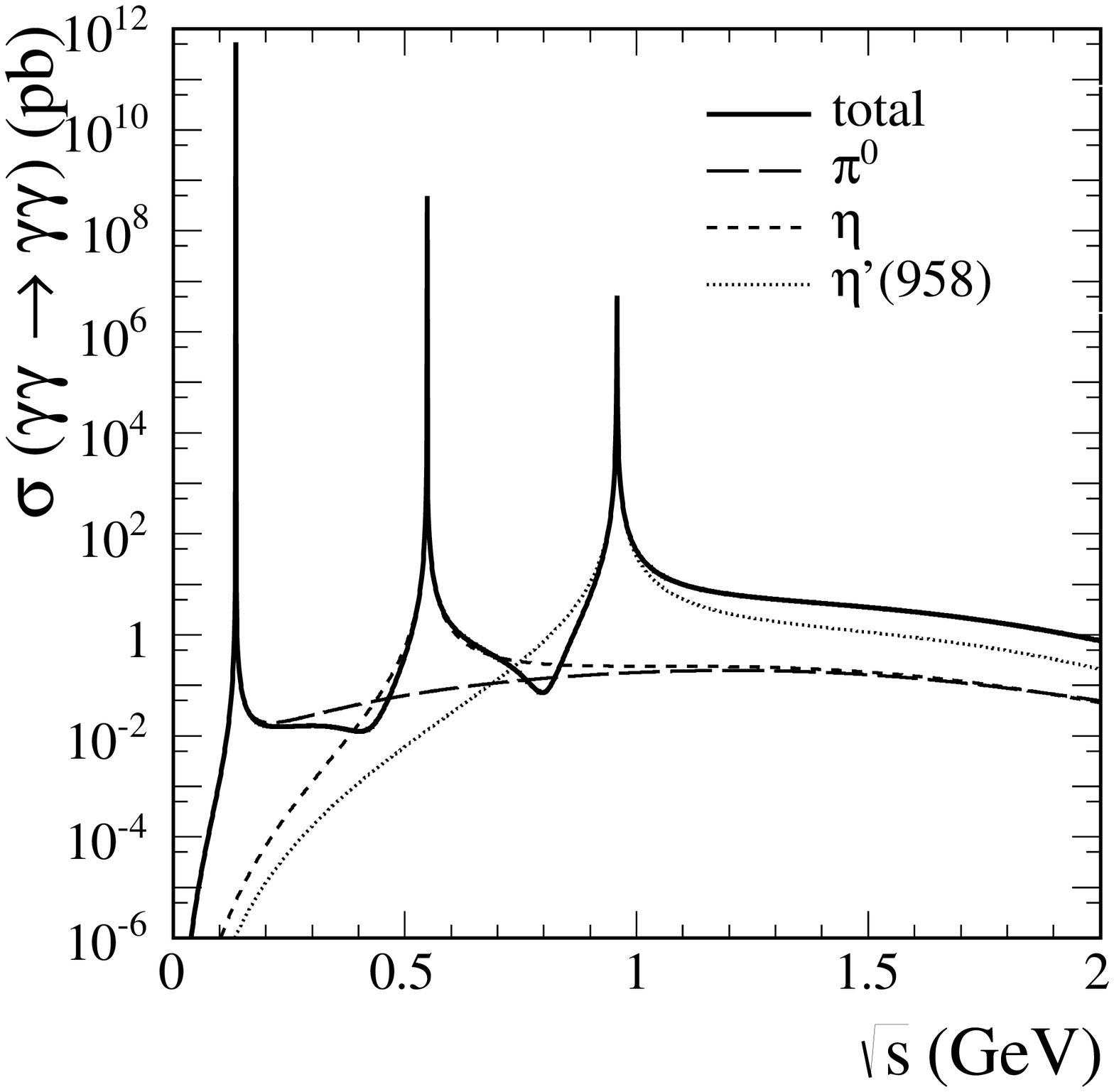}          
\label{fig:AA_PS_AA}
\caption{
The energy dependence of the pseudoscalar meson contributions
to the $\gamma \gamma \to \gamma \gamma$ scattering.
}
\end{figure}

In Fig.~3
we present an example of angular distribution at $\sqrt{s} = 1.3$~GeV
for the light pseudoscalar mesons
(similar distribution for the same energy will be discussed for 
tensor meson contributions).
The resulting distribution is relatively flat which is caused
by the dominance of the $s$-channel exchange
at the selected energy (see the previous figure). 
\begin{figure}
\includegraphics[width=8cm]{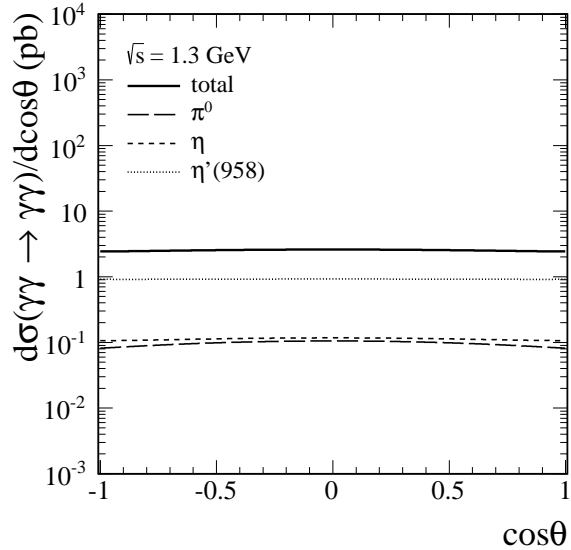}             
\label{fig:dsig_dz_PS}
\caption{
Differential cross section for the $\gamma \gamma \to \gamma \gamma$ process 
at $\sqrt{s} =1.3$~GeV.
We present results for separate pseudoscalar contributions
as well as their coherent sum.
}
\end{figure}


In the left panel of Fig.~4
we show separate contributions of tensor meson exchanges 
($s$-channel only)
as well as their coherent sum.
Similarly as for the pseudoscalar meson exchange contributions 
relatively large interference effects can be seen. 
In the right panel we show angular distributions for $\sqrt{s} = 1.3$~GeV. 
The resulting tensor meson distribution shows an enhancement
at $\cos\theta \approx \pm 1$ (compare Fig.~3 for pseudoscalar meson contributions).
The subdominant (helicity-0) $f_2(1270) \gamma \gamma$ coupling
($\Gamma^{(0)}$ in (\ref{tensor_functions})) 
plays important role for $\cos\theta \approx 0$.
\begin{figure}
\includegraphics[width=8cm]{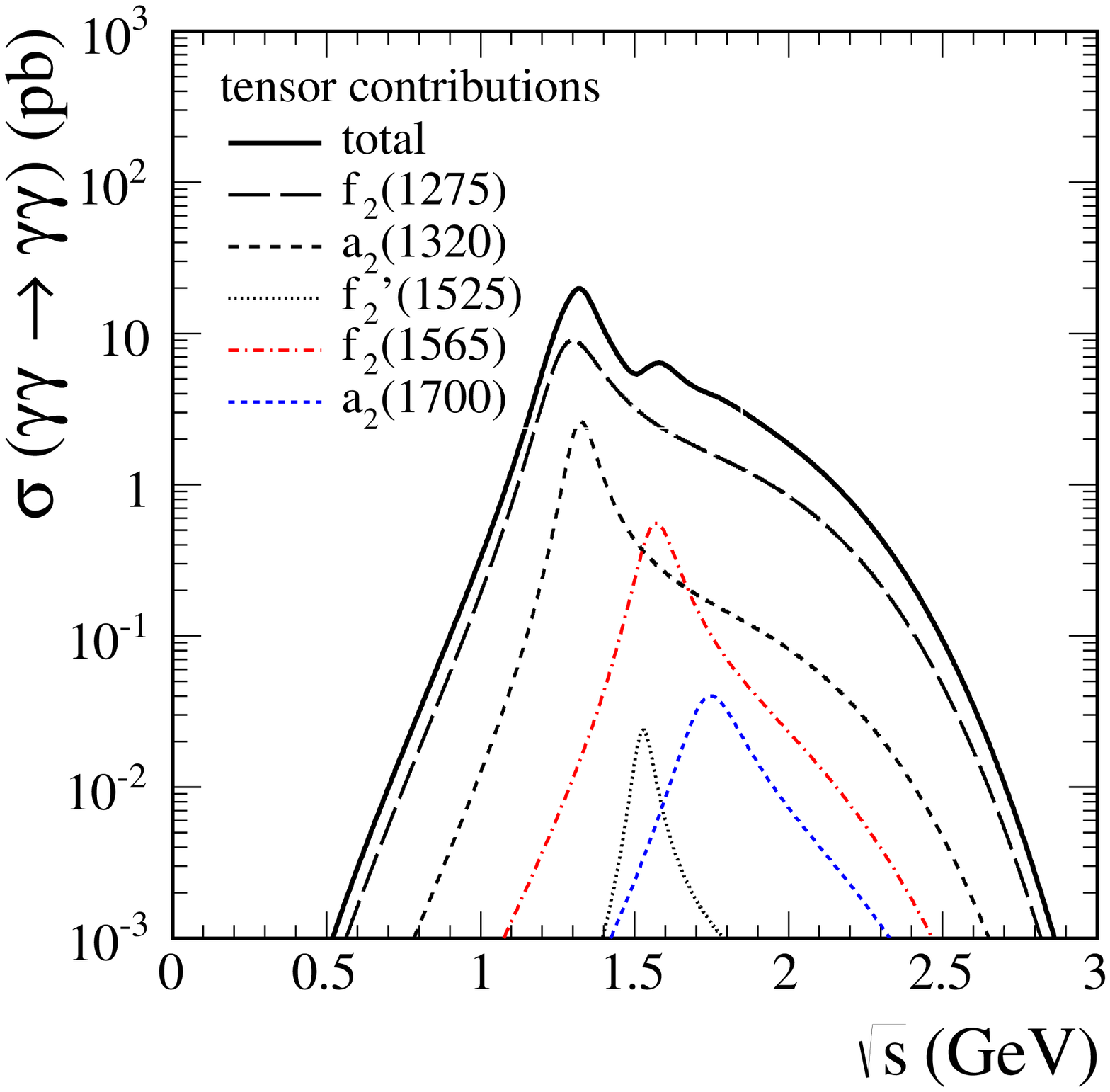}   
\includegraphics[width=8cm]{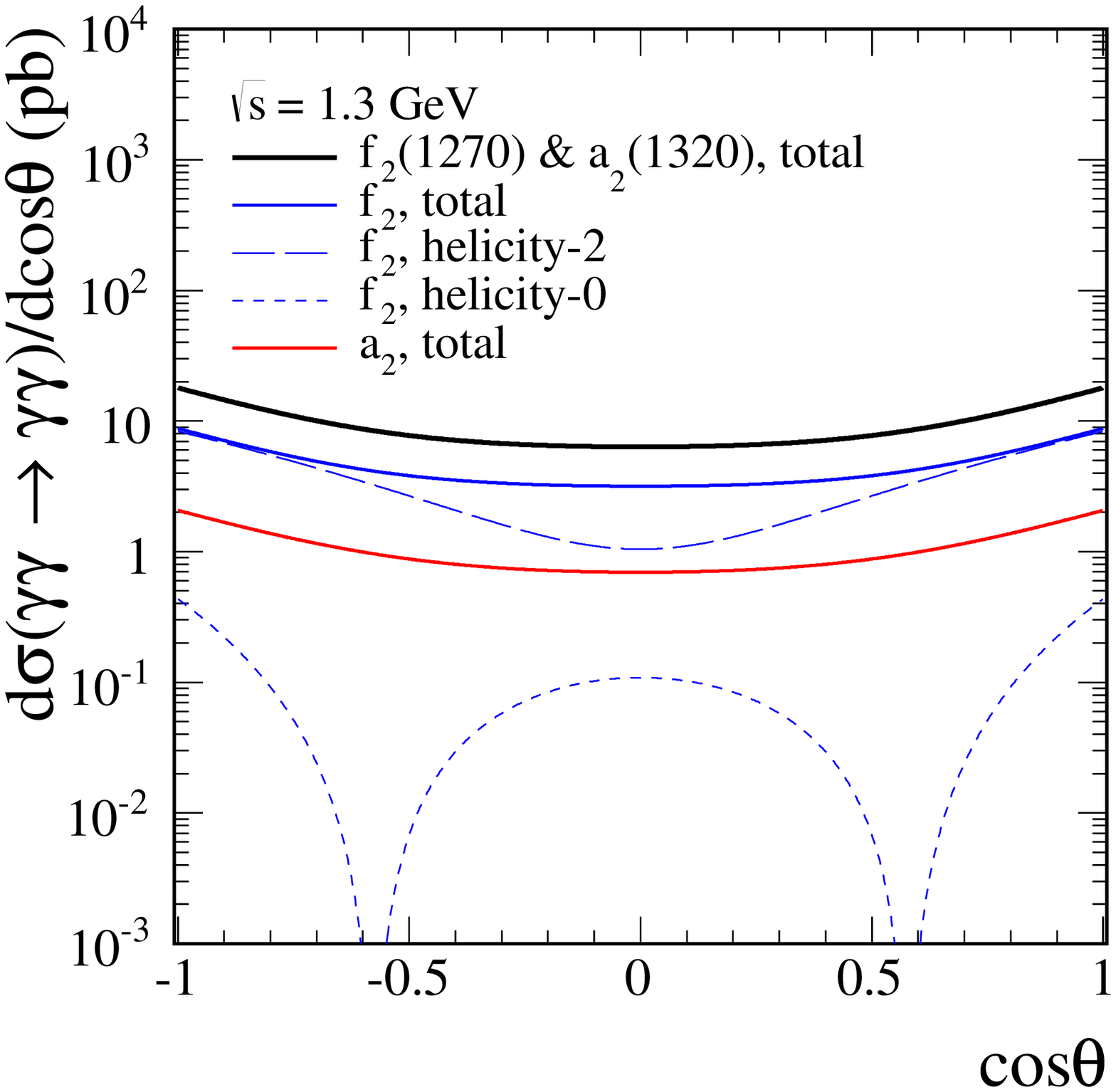}       
  \label{fig:AA_T_AA}
  \caption{
The energy dependence of the $s$-channel tensor-meson resonances (left) 
and the angular distributions at $\sqrt{s} = 1.3$~GeV (right).
}
\end{figure}

Finally, in Fig.~5
we show energy dependence 
of all meson exchange contributions 
integrated over full $z=\cos\theta$ range (left panel)
and for $|z|<0.6$ (right panel). 
For reference we show also the fermion-box (lepton and quark) 
contributions separately and added coherently together. 
The $W$-boson loops contribution plays a role only for $\sqrt{s} > 200$~GeV \cite{LPS14}.
The mesonic contribution exceed the quark-box one 
for different intervals of $\sqrt{s}$.
Only a few narrow resonances clearly exceed the total fermion continuum.
The reader is asked to note a sizeable interference effect between
pseudoscalar and tensor meson contributions.
\begin{figure}
\includegraphics[width=8cm]{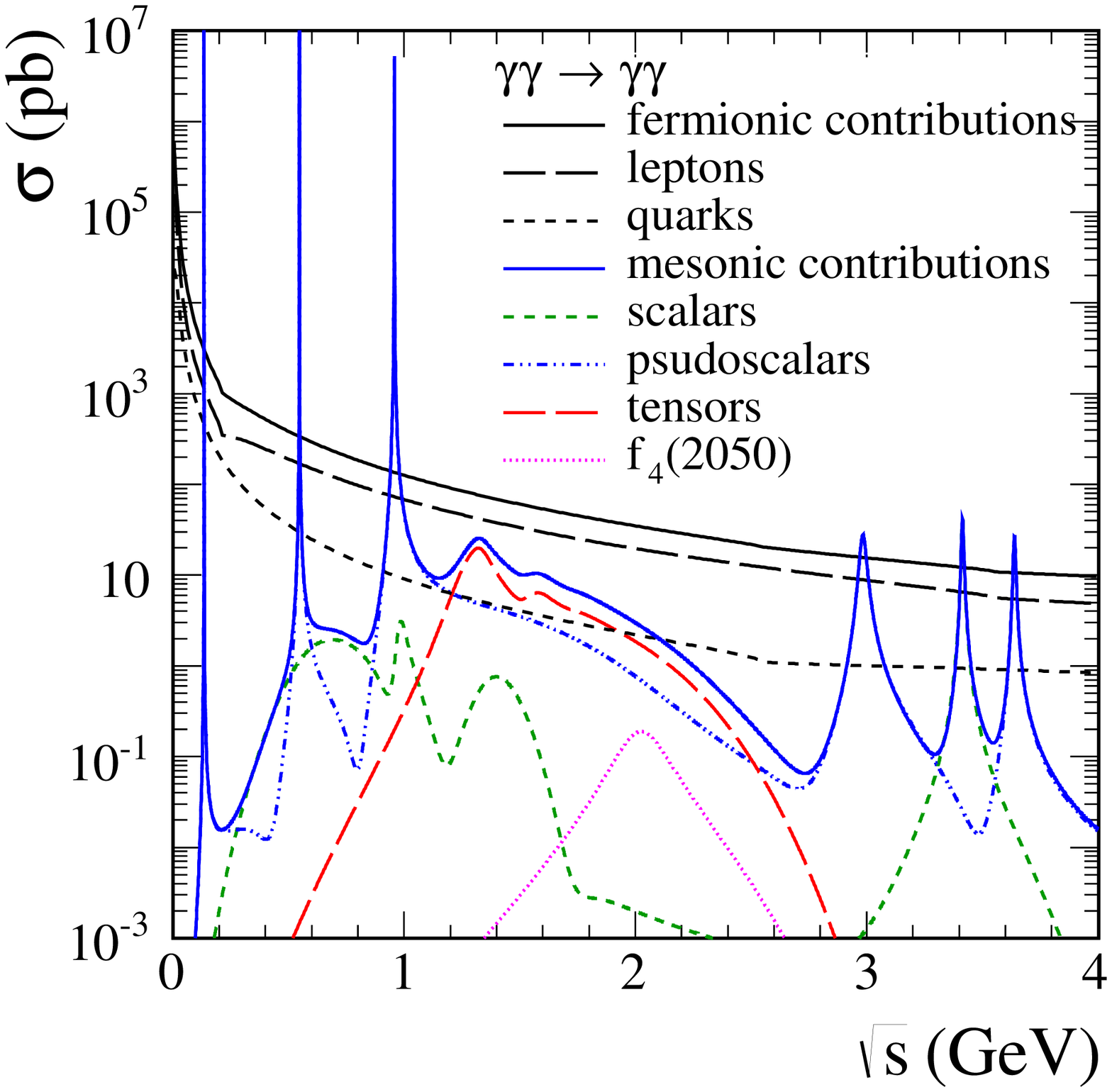}  
\includegraphics[width=8cm]{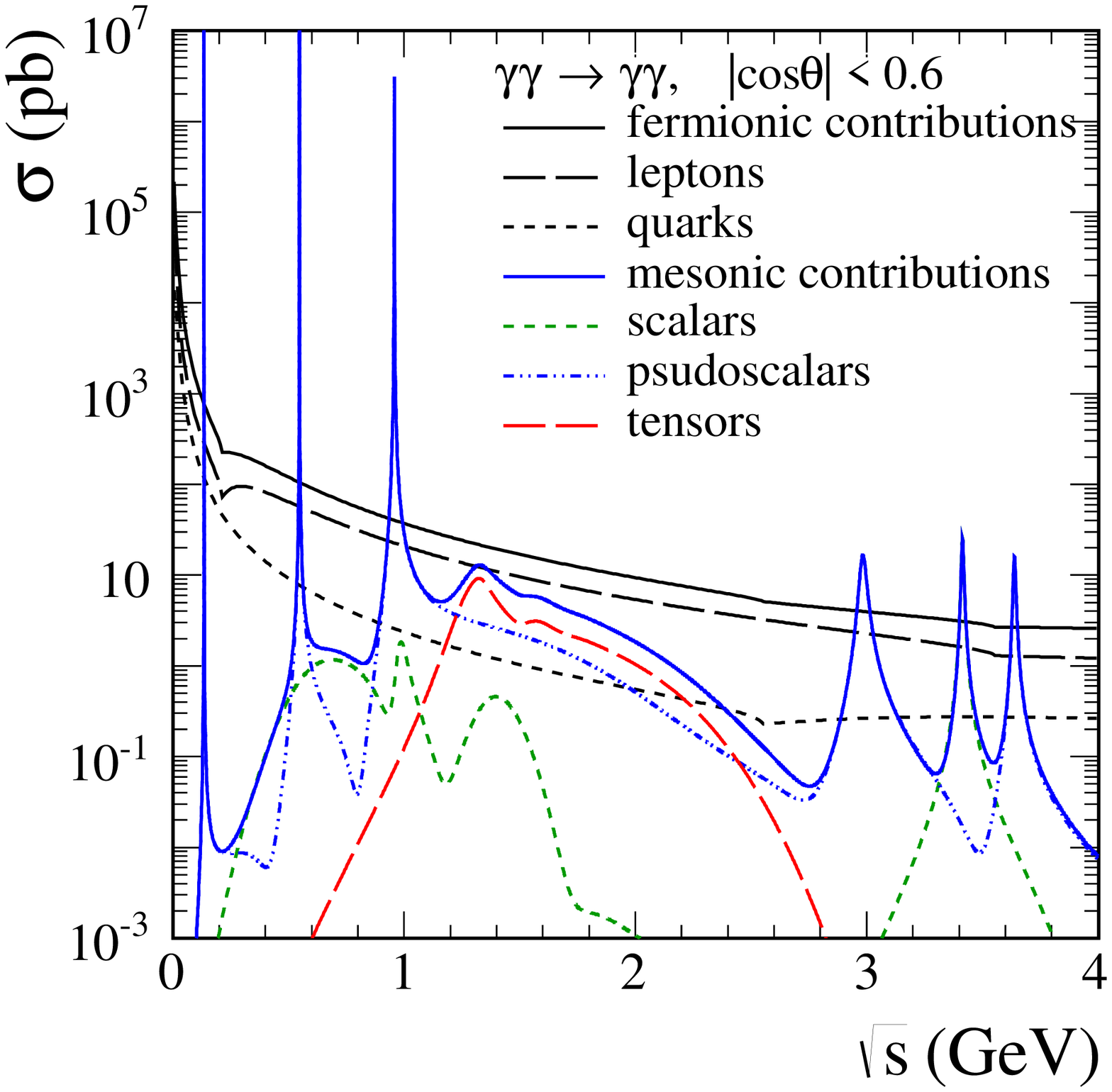}        
  \label{fig:sigma_W_all}
  \caption{
The energy dependence of the meson exchange contributions 
compared with the fermion-box ones.
Results integrated over full $z$-range (left) and for $|z|<0.6$ (right) are plotted.
The $f_{4}(2050)$ meson contribution is calculated from (\ref{sigma_f4}).
}
\end{figure}

The $\pi^0$, $\eta$, $\eta'(958)$ states were already observed
in the $e^+ e^-$ collisions. It seems rather difficult to observe
the heavy quarkonia $\eta_c(1S)$, $\eta_c(2S)$ and $\chi_{c0}$.
In heavy ion ultraperipheral collisions
at CERN this is because of poor two-photon 
invariant mass resolution of the order of 0.5 GeV while in
low-energy $e^+ e^-$ collisions because of limited phase space
and the presence of two-photon bremsstrahlung background.
The region of $f_2(1270)$ seems quite interesting as here some
enhancement could be potentially identified by the Belle II at SuperKEKB for instance.
Imposing a cut $|z|<0.6$ (see the right panel of Fig.~5) improves 
the signal (meson exchanges) to background (boxes) ratio.

The meson exchange contributions are limited
only to $\sqrt{s} < 4$~GeV, and should not influence 
the recent ATLAS experimental result \cite{Aaboud:2017bwk}.

\section{Conclusions}

In the present work we have discussed for the first time the role 
of $s$-, $t$-, $u$-channel meson exchanges in elastic 
$\gamma \gamma \to \gamma \gamma$ scattering.
We have included a few mesonic states with large two-photon branching fractions,
see Table~\ref{table:PDG}.
Large interference effects for light pseudoscalar and tensor
mesons have been found. The interference of $\pi^0$, $\eta$ and $\eta'$ amplitudes
increases the cross section for $\sqrt{s} > m_{\eta'}$ by almost an order of magnitude.
The interference of $f_2(1270)$, $a_2(1320)$ and the other tensor mesons 
leads to interesting spectral shape for collision energies in the window between 1 -- 2~GeV. 
Could this be measured at SuperKEKB in a future? Not excluded,
provided two-photon bremsstrahlung eliminated experimentally or 
is taken into account in the calculations.

We have included not only the $s$-channel diagrams 
(leading to peaks at $\sqrt{s} \simeq m_{M}$) but 
also the meson exchanges in $t$- and $u$-channels (leading to broad continua).
In general, the $t$ and $u$ diagrams contribute above the resonance peaks 
associated with the $s$-channel exchanges
which is caused by kinematics of the process and gauge invariance 
explicitly fulfilled (imposed) in our calculation.
This observation is true for the pseudoscalar, scalar and 
tensor meson exchanges.
Only the $s$-channel contributions play a role for 
heavy pseudoscalar ($\eta_c(1S)$, $\eta_c(2S)$) and scalar ($\chi_{c0}$) mesons 
and the $t/u$-channel contributions may be safely neglected.

The results related to meson exchanges have been compared with 
the standard, in the context of $\gamma \gamma \to \gamma \gamma$ scattering, 
fermion-box continuum. 
The mesonic contributions concentrate in
the region of $\sqrt{s} \in (0.1, 4.0)$~GeV and
are in general smaller than box contributions, 
except of some specific regions of the phase space.
For instance, at the resonance positions the meson exchange contributions 
sometimes even strongly exceed the standard box ones but experimental observation
may depend on diphoton invariant mass ($M_{\gamma \gamma}$) resolution of a given experiment.

The exchanges of light pseudoscalar ($\pi^0$, $\eta$ and $\eta'$)
and tensor mesons
is also important in the context of possible interference with the fermion-box contributions. 
This goes, however, beyond the scope of 
the present paper and  will be studied elsewhere.

The meson exchange contributions discussed in the present paper 
could potentially influence
the $^{208}Pb + ^{208}Pb \to ^{208}Pb + ^{208}Pb + \gamma \gamma$ 
ultraperipheral collisions measured recently by 
the ATLAS collaboration \cite{Aaboud:2017bwk}.
Our present analysis suggest, however, 
that the meson exchange contributions do not play important role 
for the recent ATLAS measurement where $M_{\gamma \gamma} > 6$~GeV.

\acknowledgments

We are indebted to Sadaharu Uehara for a discussion on the Belle
measurements of mesons in the $\gamma \gamma$ channel.
This research was partially supported by
the Polish Ministry of Science and Higher Education
Grant No. IP2014~025173 (Iuventus Plus),
the Polish National Science Centre Grant No. DEC-2014/15/B/ST2/02528 (OPUS)
and by the Center for Innovation and Transfer of Natural Sciences 
and Engineering Knowledge in Rzesz\'ow.




\end{document}